\documentclass[crop, onecolumn]{mandm}

\usepackage{setspace}

\usepackage{amsmath}
\usepackage{amssymb}
\usepackage{color}
\usepackage{graphicx}

\usepackage{caption}

\jourvolume{XX}
\jourissue{Y}
\jourpubyear{20ZZ}

\begin{document}

\title
[
Fast improvement by deep learning
]
{
Fast improvement of TEM image with low-dose electrons by deep learning
}

\author{Hiroyasu Katsuno$^{1}$, Yuki Kimura$^{1}$, Tomoya Yamazaki$^{1}$ and Ichigaku Takigawa$^{2,3}$}
\affiliation{
 $^1$Institute of Low Temperature Science, Hokkaido University, Kita-19, Nishi-8, Kita-ku, Sapporo, Hokkaido, 060-0819, Japan\\
 $^2$RIKEN, Center for Advanced Intelligence Project, 1-4-1 Nihonbashi, Chuo-Ku, Tokyo 103-0027, Japan\\
 $^3$Institute for Chemical Reaction Design and Discovery (WPI-ICReDD), Hokkaido University, N21 W10, Kita-ku, Sapporo, Hokkaido, 001-0021, Japan\\
 Corresponding Author: Hiroyasu Katsuno \email{katsuno@lowtem.hokudai.ac.jp}
}

\begin{frontmatter}

\maketitle

\begin{abstract}
Low-electron-dose observation is indispensable for observing various samples using a transmission electron microscope;
 consequently, image processing has been used to improve transmission electron microscopy (TEM) images.
To apply such image processing to \textit{in situ} observations,
 we here apply a convolutional neural network to TEM imaging.
Using a dataset
 that includes short-exposure images and long-exposure images,
 we develop a pipeline for processed short-exposure images, based on end-to-end training.
The quality of images acquired with a total dose of approximately $5$ $e^{-}$ per pixel becomes comparable to that of images acquired with a total dose of approximately $1000$ $e^{-}$ per pixel.
Because the conversion time is approximately 8 ms, \textit{in situ} observation at 125 fps is possible.
This imaging technique enables \textit{in situ} observation of electron-beam-sensitive specimens.

\noindent\textbf{Key Words:} deep learning, transmission electron microscopy, fast improvement, nanoparticles

\noindent(Received XX Y 20ZZ; revised XX Y 20ZZ; accepted XX Y 20ZZ)

\end{abstract}

\end{frontmatter}


\section{Introduction}
\label{sec:intro}

Transmission electron microscopy (TEM) is a powerful tool in the field of materials science and provides structural information through atomic-level visualization \citep{Kisielowski-2008,Morishita-MAM}.
To obtain clearer images, researchers have implemented hardware improvements in both the columns [e.g., by introducing an aberration corrector \citep{Haider-Nature1998} and a phase plate \citep{Danev-UM2001}] and cameras \citep{Faruqi-Ultramicrocopy2003} incorporated into transmission electron microscopes.
TEM has become a useful technique in various fields such as biology, electrochemistry, fluids, geology, and the environmental sciences 
 because various sample holders that enable control of the sample environments have been developed.
One of the current limitations for the further application of TEM is the influence of electron irradiation on a sample:
 A beam-sensitive material will often lose its characteristic structure under electron irradiation before an image can be acquired.
Cryo-TEM observation is an outstanding technique to reduce the electron-induced damage on a sample \citep{Moran-Science1952,Chlanda-rev-MMB2014}. 
Nevertheless,
 an additional technique is needed for low-dose observation of various samples and for \textit{in situ} imaging.
In particular, \textit{in situ} imaging using liquid-cell TEM requires both low-dose observation and high temporal resolution to reduce the beam effect,
 which is the radiolysis of the solution sample \citep{Schneider-JCCP2014}.

Possible low-electron-dose TEM imaging methods include the dictionary learning method based on sparse coding \citep{Elad-IEEE2006}.
Achieving the desired improvement requires the construction of an appropriate dictionary.
When an appropriate dictionary is used,
 a fragment of the image is replaced by a linear combination of basic elements.
Sparse coding has been successfully used to improve electron tomography images \citep{Binev-EMNST}, scanning TEM imaging \citep{Stevens-Mcro2014}, and electron holography \citep{Anada-hologram}.

Over the past decade,
 image taken by general cameras have been dramatically improved by machine learning.
In addition to sparse coding,
 regarding image denoising,
 numerous methods have been proposed for image denoising, including sub-pixel convolutional neural network (CNNs) \citep{Shi-arxiv2016}, nuclear norm minimization \citep{Gu-IEEE2014}, and domain filtering \citep{Dabov-IEEE2007}.
These techniques are based on topics such as smoothness and/or sparsity and useful for still images.
However, these techniques have a disadvantage for \textit{in site} observations because of computational cost.
For example, the time necessary for image improvement via sparse cording is about 1--10 s \citep{Anderson-2013}.

Recently, 
 a new technique of deep learning has been proposed for low-light image enhancement.
In this method,
 an images is acquired with a low-light image so that the original image can be used as ground truth for comparison \citep{Lim-ICIP2015}.
A CNN is introduced, and an image set of short-exposure and long-exposure images are prepared for training \citep{Chen-SID}.
The corresponding results have shown that not only the machine learning model but also datasets need improvement.

In general, image conversion is fast, although it takes a long time to train a CNN model.
Using a CNN may help increase the speed of image improvement, enabling \textit{in situ} observation.
In the present work, we apply a CNN model to TEM imaging and evaluate the image quality and the speed for image improvement.

\section{Methods}
\label{sec:methods}
\begin{table}
\centering
\caption{The number of images for training and validation.}
\begin{tabular}{ccccc}
\hline
\ \ \ Set No.\ \ \  & \ \ \ Material \ \ \ & \ \ \ Training\ \ \  & \ \ \ Validation \ \ \   \\
\hline
1 & Ni              & 224 & 36 \\
2 & Fe--Ni          & 260 & 40 \\
3 & SiC             & 140 & 20 \\
4 & Silicate        & 176 & 24 \\
5 & Alumina         & 200 & 40 \\
\hline
\end{tabular}
\label{tab:numofimages}
\end{table}

\begin{figure*}
\centering
\includegraphics[width=\textwidth, bb = 0 200 960 380, clip]{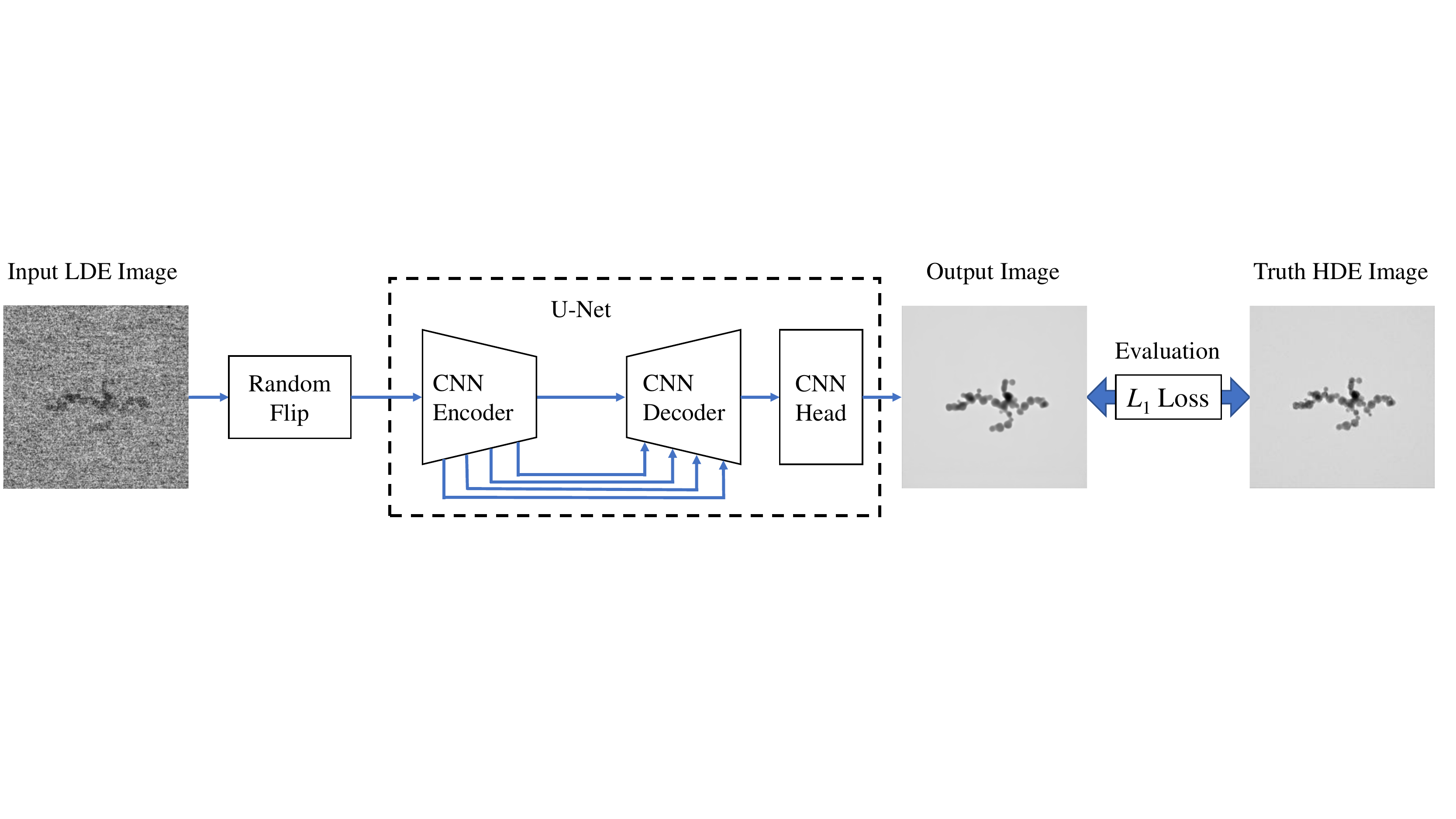}
\caption{
Schematic of image improvement using our deep learning model.
}
\label{fig:ourmodel}
\end{figure*}

We used a transmission electron microscope equipped with a field-emission gun (JEM-2100F, JEOL, Tokyo)
 operated at an acceleration voltage of 200 kV and a CMOS camera, OneView IS (Gatan, Inc., Pleasanton, CA, USA).
For training, we prepare sets of high-dose-electron (HDE) and low-dose-electron (LDE) TEM images with the same field of view.
All images were acquired with drift correction using the function incorporated into the software (Digital Micrograph, Gatan. Inc., Pleasanton, CA, USA) used to operate the CMOS camera.
The HDE image resolution was $4096 \times 4096$ pixels, and its exposure time was $5$ s.
After an HDE image was acquired, the corresponding LDE image is taken at a resolution of $512 \times 512$ pixels and with an exposure time of $3.3$ ms.
The typical total doses in each view were $10^{10}$ $e^{-}$ and $10^{6}$ $e^{-}$ in HDE and LDE images, respectively, on the camera.
The dose rate was calibrated using a Faraday cage (JEOL Ltd., Tokyo, Japan).
The numbers of electrons on each pixel were $\sim 1000$ $e^{-}$ and $\sim 5$ -- $10$ $e^{-}$ in the HDE and LDE images, respectively.
Typical magnifications used in the present study were $25,000\times $  and $30,000\times$; the dose rate on the samples was therefore $\sim 10^{2}$ $e^{-}$nm$^{-2}$s$^{-1}$ in case of LDE observation.
The number of images for training data is summarized in Table~\ref{tab:numofimages}.
The typical samples were particles of Ni, FeNi alloy, SiC, silicate, and alumina with diameters of $30$--$200$ nm.

The original binary data of the digital micrograph were converted to grayscale images in tiff format with the dark current subtracted and including the intensity with 16-bit expression in each pixel.
The obtained images were preprocessed with intensity rescaling.
Before training, 
 the position of each HDE image was adjusted so that its position fit the corresponding LDE image,
 thereby compensating for sample drift in the sequences.

Our deep learning model has the U-Net architecture \citep{Ron-unet} with the ResNet encoder/decoder \citep{He-resnet} using the segmentation package in PyTorch \citep{smp-pytorch}.
The schematic of our image improvment is shown in Fig.~\ref{fig:ourmodel}.
The model parameters were fine-tuned against the pre-trained Resnet-18 model from the ImageNet data
 using the $L_{1}$ loss function and the Adam optimizer \citep{Kingma-adam} with the learning rate of $10^{-4}$.
The concrete form of $L_{1}$ function for two images is written as
\begin{align}
L_{1} = \frac{1}{N}\sum_{x} |I_{\rm A}(x)-I_{\rm B}(x)|,
\label{eqn:l1}
\end{align}
where $x$ represents the position of a pixel.
$I_{\rm A}(x)$ and $I_{\rm B}(x)$ are the normalized intensity of two images at the position $x$, and $N$ is the total number of pixels.
In each iteration,
 an LDE image was randomly flipped horizontally and/or vertically for data augmentation
 and the random crop was not applied.
All training was conducted on a Linux machine with 10-core Intel i9-9900X 3.5 GHz CPU and an NVIDIA Quadro RTX 8000 graphics card (see Machine No. 1 in Table~\ref{tab:wt}).

\section{Results \& Discussion}
\label{sec:results}

\subsection{Overview of the Image Improvement}

\begin{figure}
\centering
\includegraphics[width=0.5\textwidth]{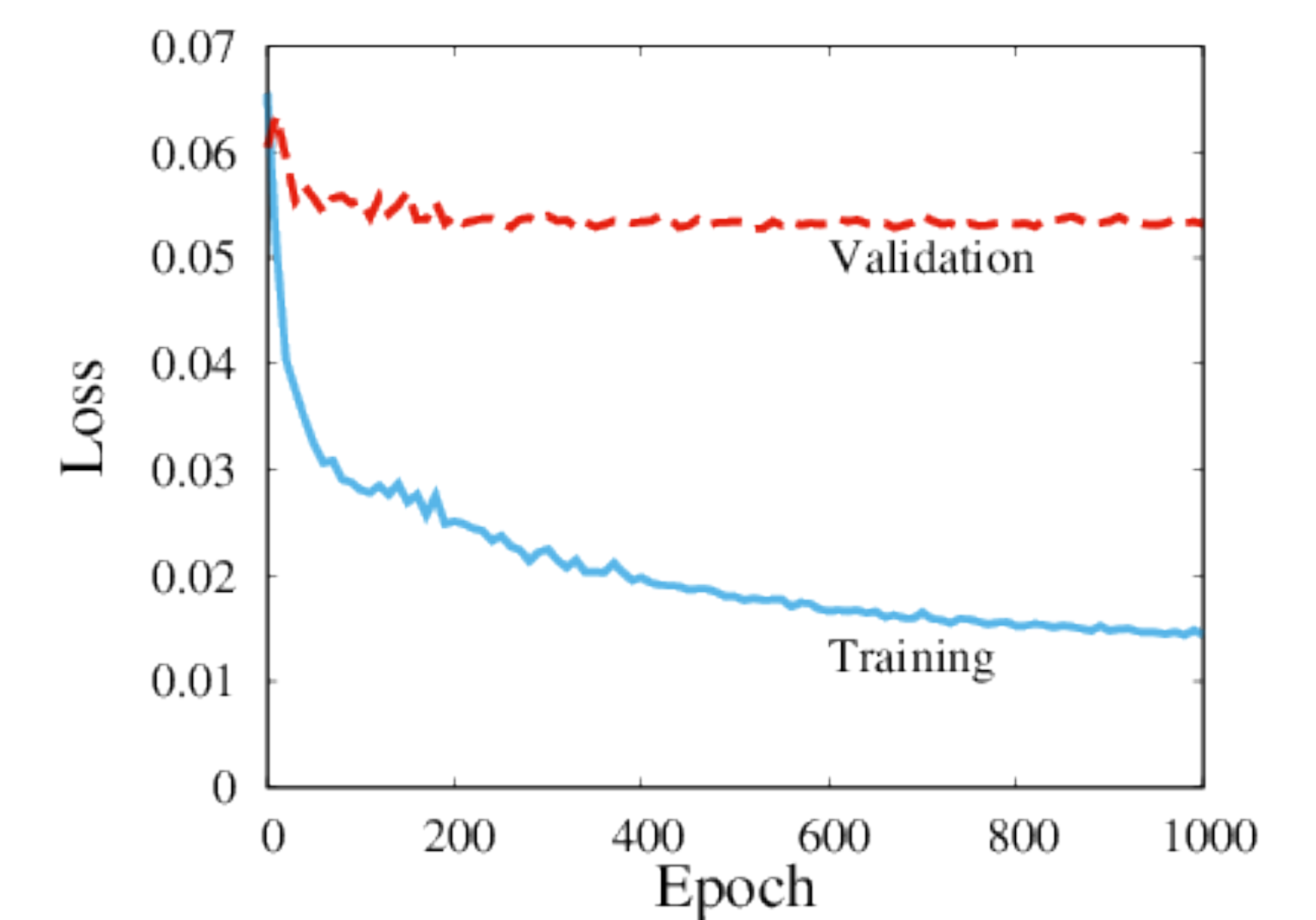}
\caption{Training loss and validation loss.}
\label{fig:lossfigs}
\end{figure}

\begin{figure*}
\centering
\includegraphics[width=\textwidth, bb = 0 190 960 325, clip]{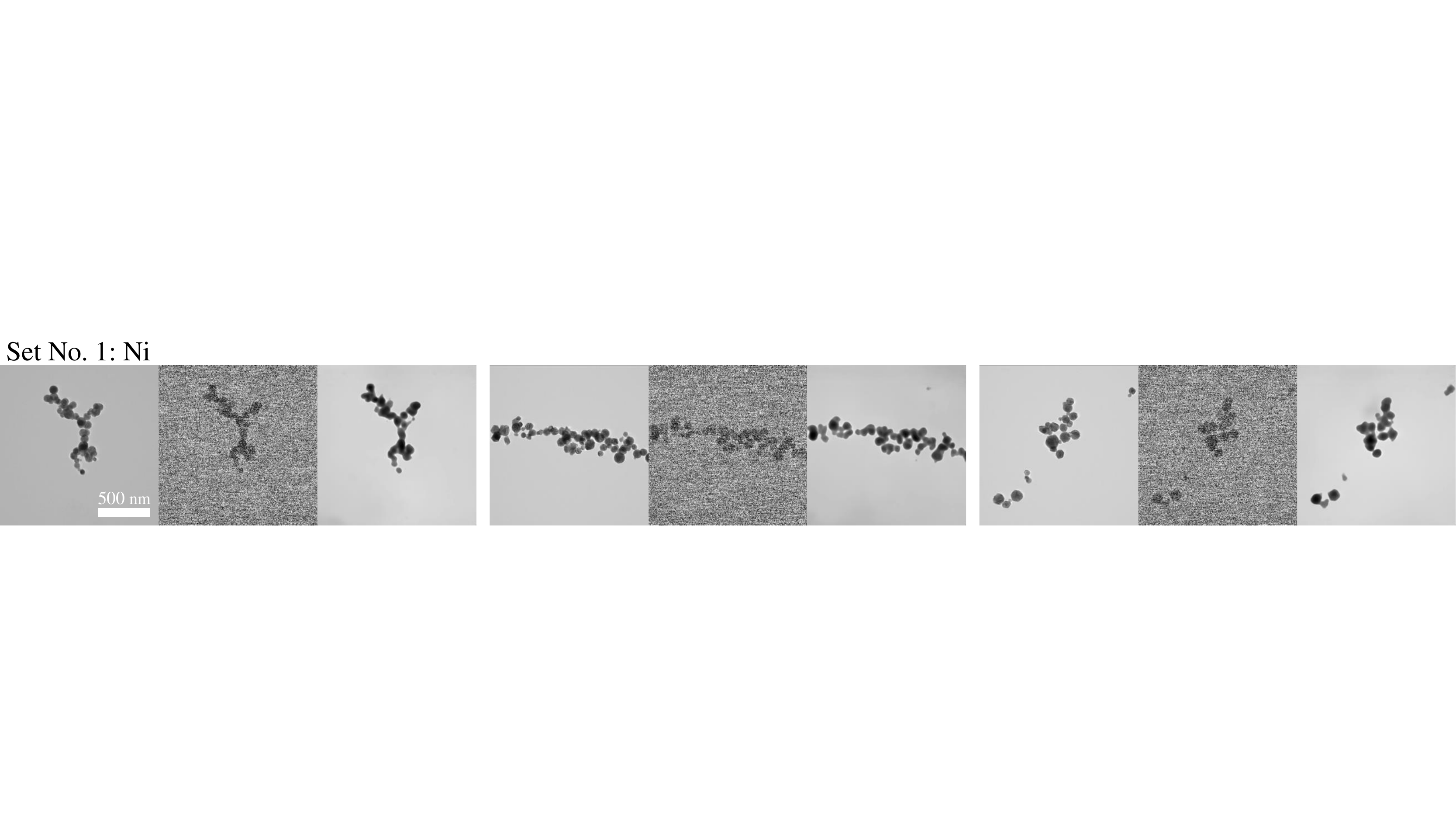}
\includegraphics[width=\textwidth, bb = 0 190 960 325, clip]{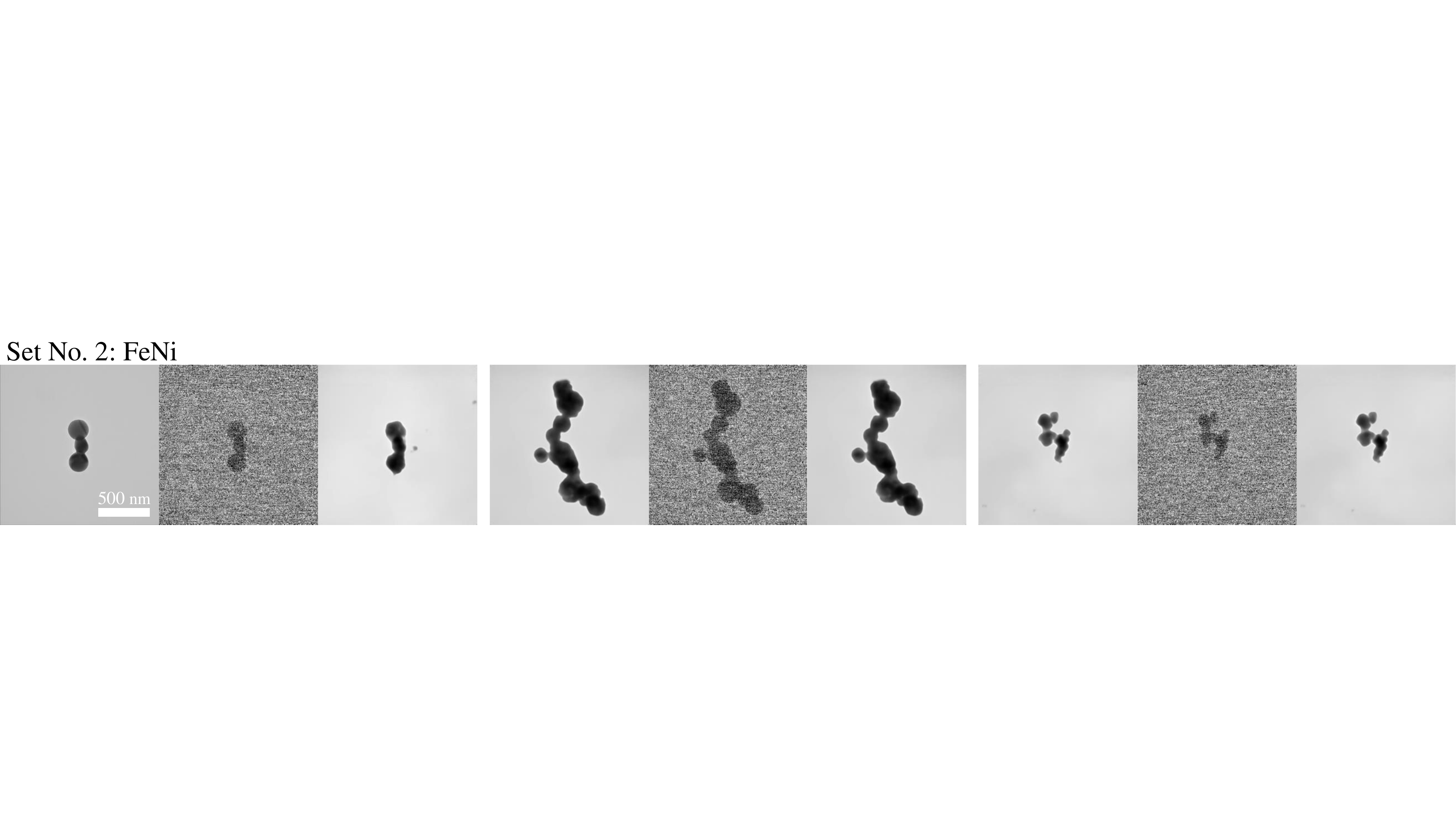}
\includegraphics[width=\textwidth, bb = 0 190 960 325, clip]{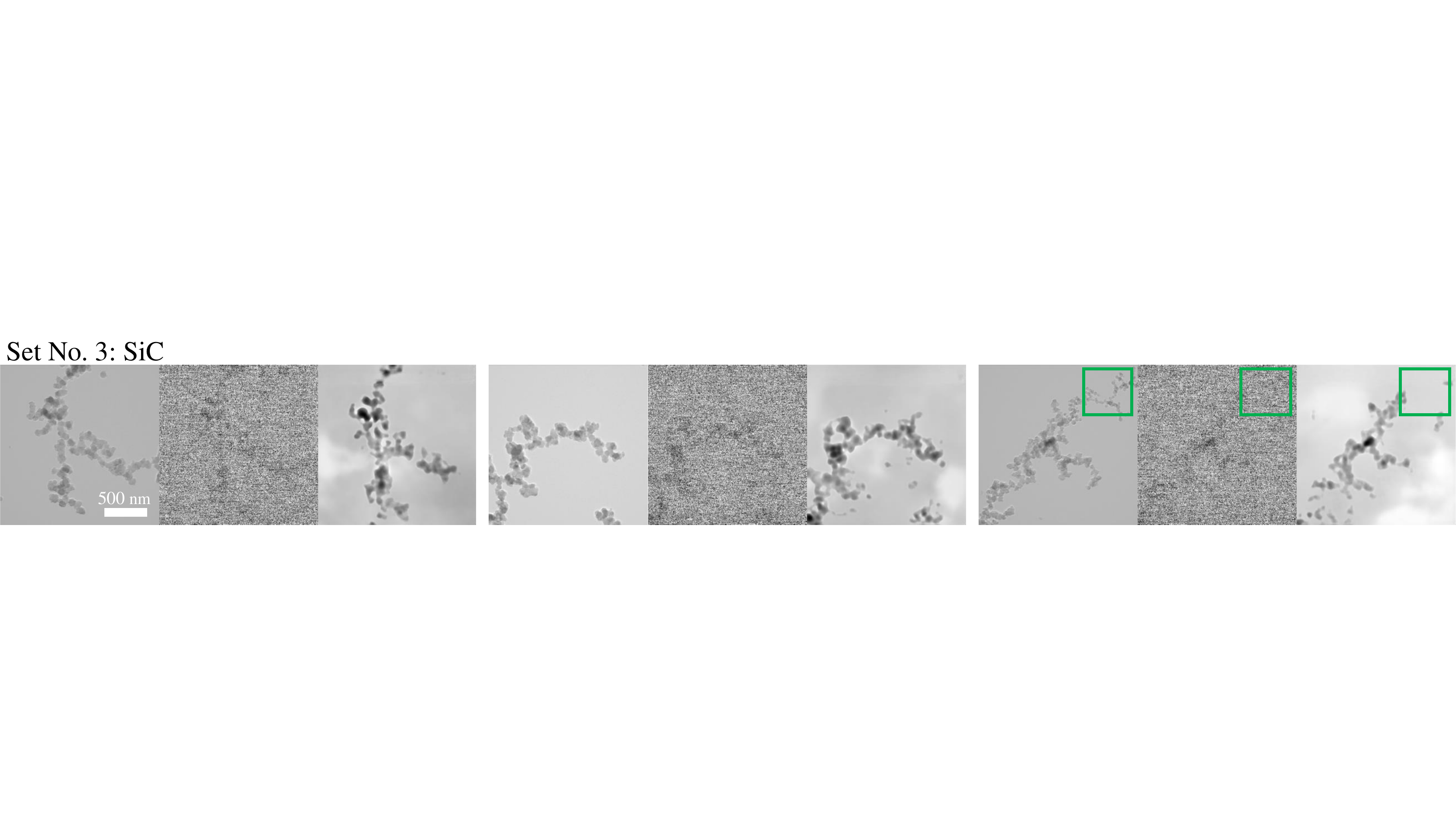}
\includegraphics[width=\textwidth, bb = 0 190 960 325, clip]{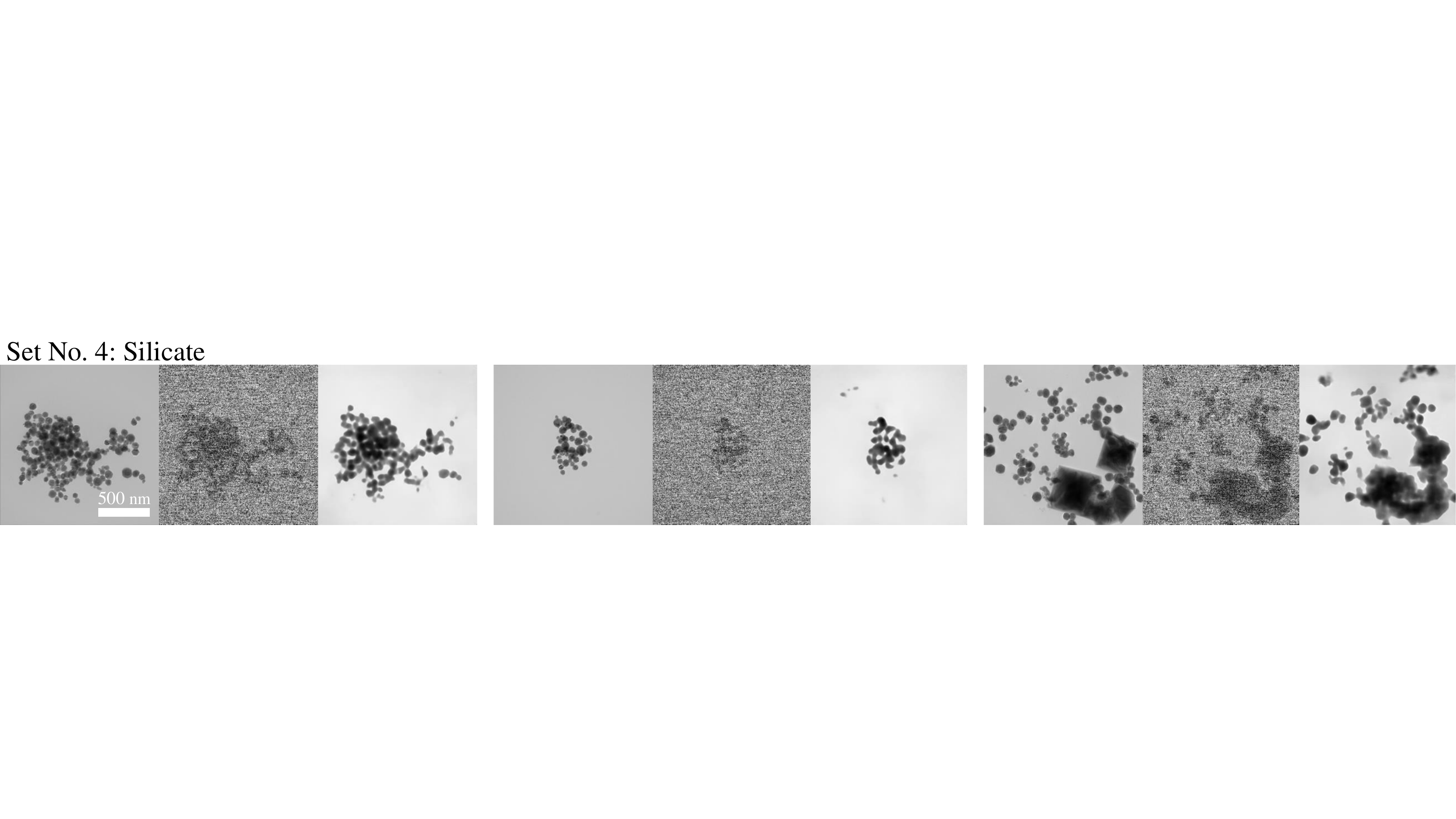}
\includegraphics[width=\textwidth, bb = 0 190 960 325, clip]{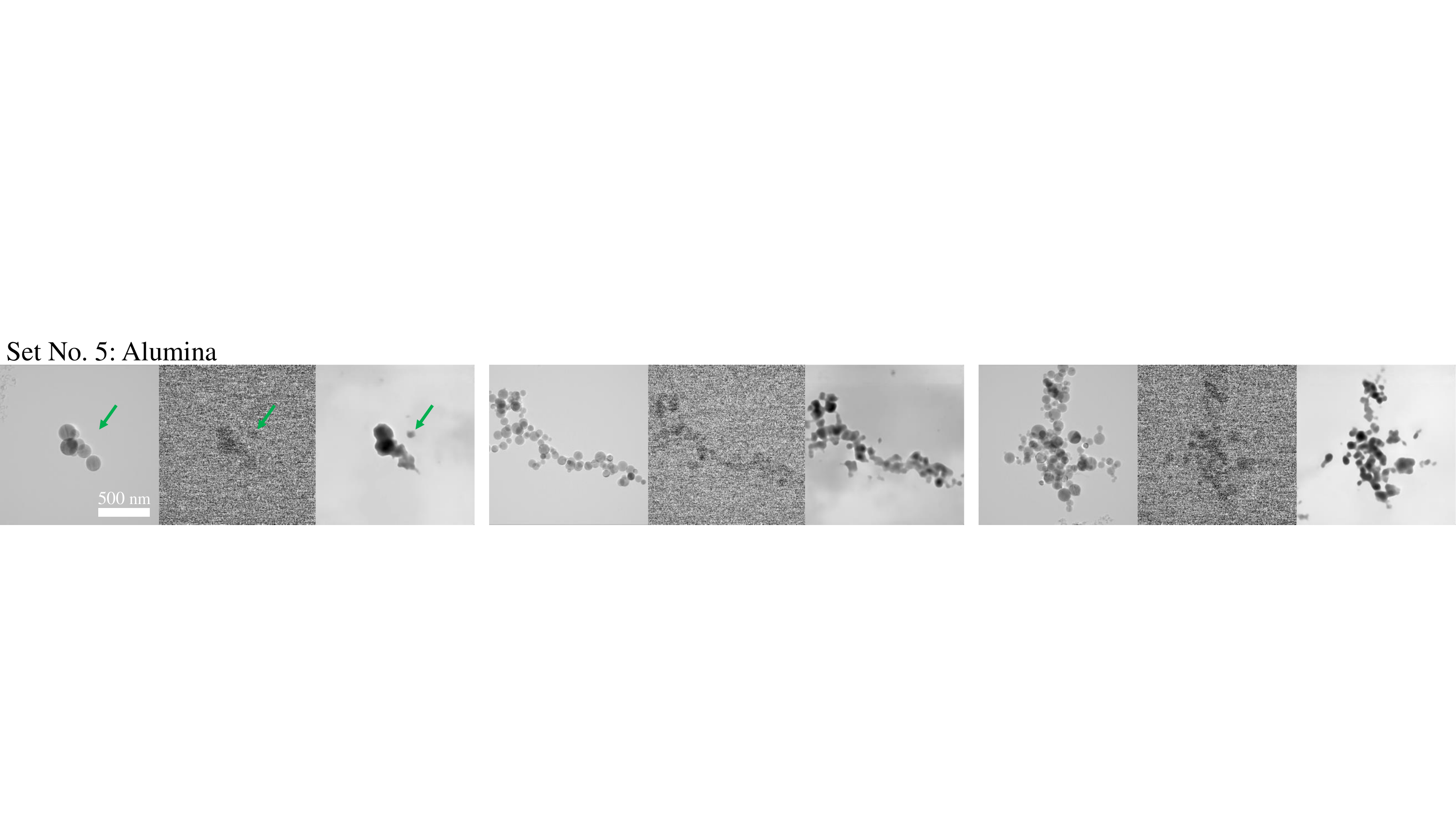}
\caption{
Examples of HDE, LDE and output images (left to right); images of Ni, FeNi, SiC, silicate, and alumina displayed top to bottom.
Arrows and boxes indicate examples of the failure results of the improvement.
The presented images were used for validation, not training.
}
\label{fig:examples}
\end{figure*}

\begin{table*}
\centering
\caption{Qualitative comparison of output images for validation using the mean absolute error (MAE) and the peak signal-to-noise ratio (PSNR).}
\begin{tabular}{ccccc}
\hline
\ \ \ Set no.\ \ \  &\ \ \ Material \ \ \ &\hspace*{3em} MAE \hspace*{3em} & \hspace*{3em} PSNR \hspace*{3em} \\
\hline
1 & Ni              & 0.075 $\pm$ 0.026 & 20.82 $\pm$ 2.13\\
2 & FeNi            & 0.034 $\pm$ 0.011 & 25.10 $\pm$ 1.86\\
3 & SiC             & 0.074 $\pm$ 0.020 & 20.80 $\pm$ 1.60\\
4 & Silicate        & 0.062 $\pm$ 0.025 & 21.60 $\pm$ 2.82\\
5 & Alumina         & 0.075 $\pm$ 0.026 & 20.69 $\pm$ 2.06\\
\hline
& (Average) & 0.063 $\pm$ 0.028& 21.97 $\pm$ 2.79\\
\hline
\end{tabular}
\label{tab:L1andPSNR}
\end{table*}

Figure \ref{fig:lossfigs} shows the epoch dependence of the loss of the training dataset and the validation dataset.
As the training proceeded, the value of the loss decreased until 100 epochs.
Whereas the loss value for the training data decreased after 100 epochs,
 that for the validation data was steady.
We also checked the loss in more than 1000 epochs,
 and observed a slight increase in the loss value of the validation data because of overfitting.
In the present study,
 we used the finetuned parameter at 900 epochs.

Figure \ref{fig:examples} shows some examples of HDE images,
 LDE images,
 and output images obtained using our model from corresponding LDE images.
These HDE and LDE image sets were not used for training.
Images of Ni, FeNi, SiC, silicate, and alumina are displayed from the top row to the bottom row in Fig.~\ref{fig:examples}.
Various sizes and contrasts of nanoparticles are observed in HDE images.
The preprocessed LDE images are similar to the images displayed by the camera software (Gatan Digital Micrograph) during low-dose-rate observations.
All of the images show noise resembling a sandstorm.
A common feature of all of the output images is that the noise is removed,
 although a haze remained in the background of the images in the third and fifth rows of Fig.~\ref{fig:examples}.
The overlapping nanoparticles, such as those in the fourth row and first column image of Fig.~\ref{fig:examples}, were reproduced in the output images.
Thus, we obtained an image comparable to an HDE image from a sandstorm image,
 although the improvement failed in some cases:
 the appearance of imaginary nanoparticles indicated by the arrow in the image in the fifth row and first column of Fig.~\ref{fig:examples},
 and the disappearance of nanoparticles, such as in the boxed area in the image in the third row and third column of Fig.~\ref{fig:examples}.

The quality of the improvement appears to depend on the sample.
For a qualitative comparison of the quality of the output images,
 the mean absolute error (MAE) and the peak signal-to-noise ratio (PSNR) are listed in Table~\ref{tab:L1andPSNR}.
The MAE is the difference between images at the pixel level (see details in the next subsection)
 and the large MAE value indicates a large difference between images.
The PSNR is a logarithm of the inverse of the mean square error
 where a large value of PSNR indicates that two images are similar.
Both indicators show that 
 our model is more effective for set No.~2 (FeNi) than the other sets.
In set No.~2,
 the magnitude of the electron count on nanoparticles is less than 50\% of that on the background in HDE images.
However,
 in sets No.~3 (SiC) and No.~5 (alumina),
 the magnitude of the electron count on the nanoparticles is almost 90\% of that on the background in HDE images;
 i.e., most of the SiC and alumina nanoparticles exhibit relatively weaker contrast than FeNi particles.
The small difference in contrast between the background and the nanoparticles, which is only 10\%,
 may prevent the image improvement.
In sets No.~1 (Ni) and No.~4 (silicate),
 the HDE images tend to aggregate nanoparticles with weak and strong contrast.
In particular, the large standard deviation of the PSNR of set No.~4 might be reflected the variety of images.
At least,
 the statistical data indicate
 that our model provides the same level of image quality for various materials.

\subsection{Example of Visualization of LDE Image}

In a microscope image,
 indicators of the performance include the accuracy of the size of an object and the accuracy of the separation of two adjacent objects.
We studied the improvement rate and these two indicators in detail as an example in set No.~4.

\begin{figure*}
\centering
\includegraphics[width=\textwidth, bb = 0 100 960 440, clip]{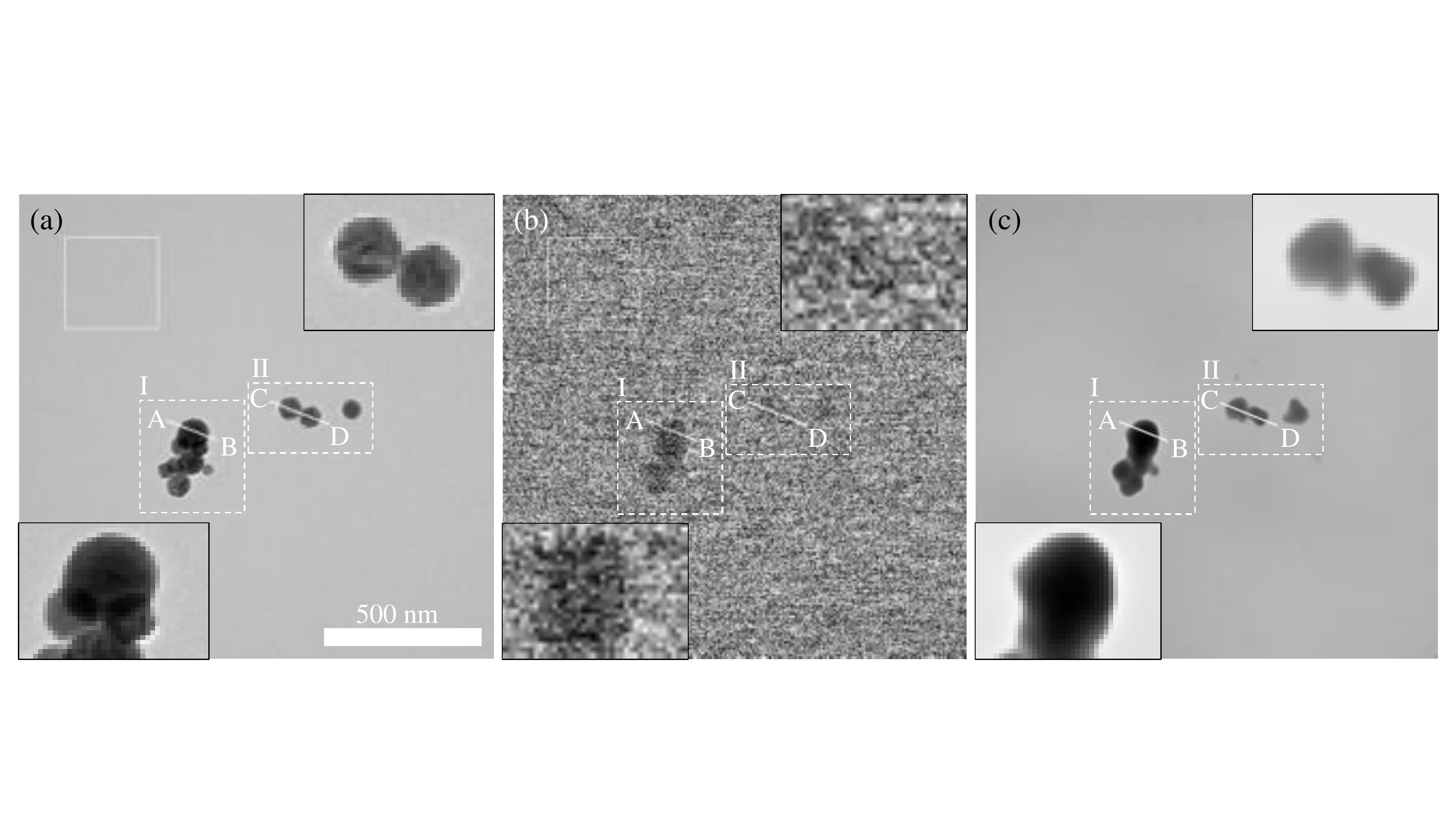}
\caption{
Examples of (a) an HDE image, (b) a LDE image and (c) an output image of set No.~4.
The average dose is about 1200 $e^{-}$ per pixel and 4.7 $e^{-}$ per pixel in a box.
The insets in the figures show enlarged images of parts of region I and II.
The presented images were used for validation, not training.
}
\label{fig:rba}
\end{figure*}

\begin{figure*}
\centering
\includegraphics[width=\textwidth, bb = 0 0 960 540, clip]{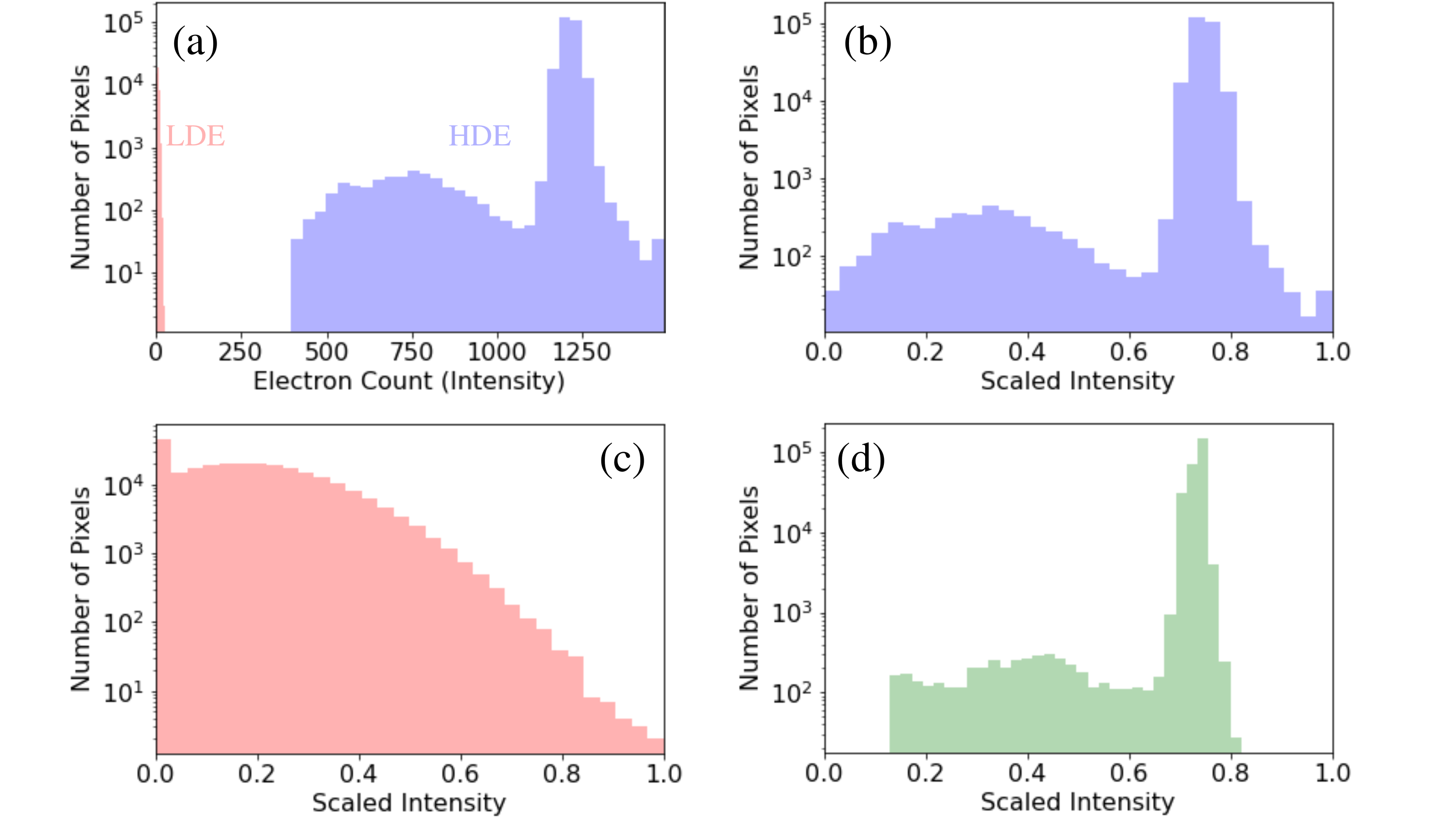}
\caption{
(a) Raw intensity histograms of the HDE and LDE images shown in Fig.~\ref{fig:rba}.
Scaled intensity histograms of (b) the HDE image, (c) the corresponding LDE image, and (d) the output image in Figs.~\ref{fig:rba}(a)--(c), respectively.
The abscissa is the electron count and the ordinate is the number of pixels.
The data is plotted as semi-log plots.
}
\label{fig:hist_p}
\end{figure*}

Figure \ref{fig:rba} shows an example of (a) an HDE image, (b) an LDE image, and (c) a corresponding output image from (b).
In Fig.~\ref{fig:rba}(a), there are nine silicate nanoparticles in region I indicated by the dashed-line box and three nanoparticles in region II.
The nanoparticles are typically 100 nm in diameter.
Figure \ref{fig:rba}(b) shows a preprocessed LDE image.
Sandstorm-like noise is present in the whole image.
Only two particles are recognized in region I,
 and it is difficult to recognize a particle in region II.
Figure \ref{fig:rba}(c) is the output image generated from Fig.~\ref{fig:rba}(b) by our model.
Particles are recognized in both regions I and II, and no particles are observed elsewhere, as in Fig.~\ref{fig:rba}(a).

All images have been normalized and have intensity values from 0 to 1 in each pixel,
 and $L_{1}$ function indicates the difference in the intensity at pixels (see Eq.~\ref{eqn:l1}).
The maximum value of $L_{1}$ is $1$, where an image is composed of only black and white pixels 
 and another image is composed of the opposite color, e.g., a set of all black and all white images. 
On the contrary, the minimum value is 0; where two images are in complete agreement.
The value of the $L_{1}$ loss in Fig.~\ref{fig:rba}(a) and Fig.~\ref{fig:rba}(b) is $0.34$.
After the training, the value decreases to $0.02$ in the case of Fig.~\ref{fig:rba}(a) and Fig.~\ref{fig:rba}(c).
The image is improved by our model.

Figure~\ref{fig:hist_p}(a) shows the raw intensity histograms of the HDE and LDE images shown in Fig.~\ref{fig:rba} as semi-log plots.
As the minimum electron count of the HDE image is 396 and the maximum value of the LDE image is 24, the histogram of the LDE image is barely visible in this plot.
The histograms of the HDE and the LDE images are scaled as shown in Figs.~\ref{fig:hist_p}(b) and \ref{fig:hist_p}(c), respectively.
On the histogram corresponding to the HDE image [Fig.~\ref{fig:hist_p}(b)] shows two peaks.
The sharp peak arises from the background, whose value is about 0.7--0.8.
Another broad peak at $0.3$ originates from the nanoparticles.
Two peaks are clearly separated on the intensity histogram.
However, the intensity histogram corresponding to the LDE image [Fig.~\ref{fig:hist_p}(c)] shows a single peak.
The intensity of the nanoparticles is not substantially different from that of the background.
Figure \ref{fig:hist_p}(d) shows the histogram of the output image.
Despite the conversion of the LDE image [Fig.~\ref{fig:rba}(b)], whose histogram shows a broad single peak [Fig.~\ref{fig:hist_p}(c)],
 a sharp peak at 0.7 and a broad peak at 0.4 appear.
The shape of the histogram is similar to that of the histogram of the HDE image
 although there is no intensity greater than 0.8 or smaller than 0.15.

\begin{figure*}
\centering
\includegraphics[width=\textwidth, bb = 40 150 900 390]{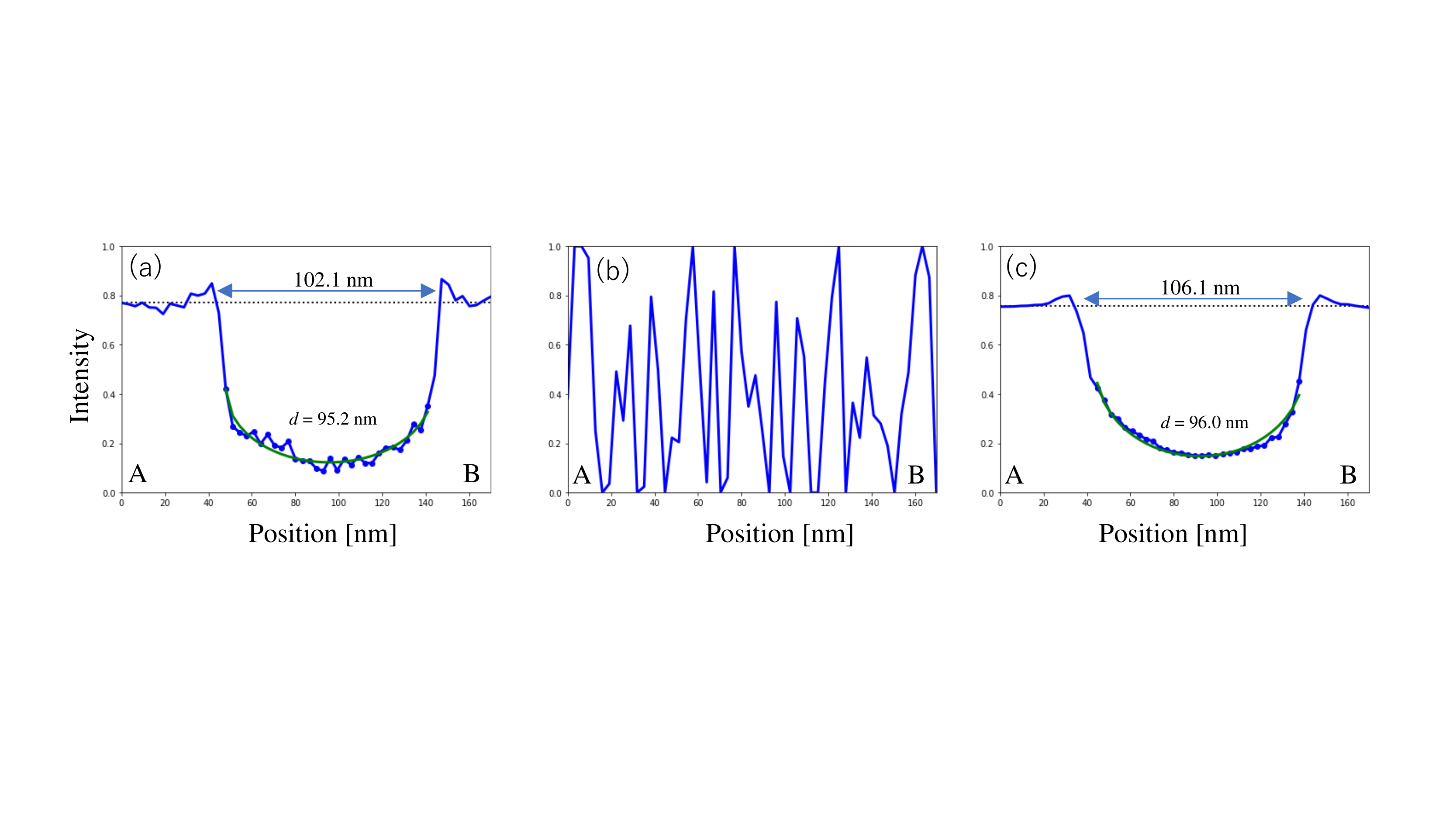}
\caption{
Line profile of A--B shown in Fig.~\ref{fig:rba}:
(a) the HDE image, (b) the LDE image, and (c) the output image.
Circles are the data used to calculate the diameter of the nanoparticle.
The dotted line shows the average of the background.
}
\label{fig:lineprofileAB}
\end{figure*}

We here focus on the discrimination of nanoparticles from viewpoint of image improvement.
We first investigated whether the size of nanoparticles could be accurately reproduced in the output images.
A magnified view of location A--B in the LDE image is shown in the bottom-left inset in Fig.~\ref{fig:rba}.
Although all nanoparticles are easily found in the HDE and output images,
 all of the nanoparticles are difficult to observe in the LDE image.
The line profiles corresponding to A--B in Fig.~\ref{fig:rba} are shown in Fig.~\ref{fig:lineprofileAB}.
In the HDE image [Fig.~\ref{fig:lineprofileAB}(a)] and the output image [Fig.~\ref{fig:lineprofileAB}(c)],
 a concave curve appears at the center because of the nanoparticle.
In the LDE image shown in Fig.~\ref{fig:lineprofileAB}(b),
 random noise is present and the concave curve does not appear.
When the shape of the nanoparticle is assumed to be spherical,
 the diameter of the nanoparticle is found to be $95.2$ nm in the HDE image and $96.0$ nm in the output image
 on the basis of the data corresponding to the center of the concave curve.
The output image generated by our model reproduces the size of the nanoparticle with an accuracy within $1$\%.

In addition, we investigated the edge width of a nanoparticle.
The edge width was estimated by comparing a region of strong contrast and the estimated diameter of a particle with an assumed spherical shape.
The distance between crosspoints composed of the line profile (blue solid line) and the background (horizontal dotted line) in Figs.~\ref{fig:lineprofileAB}(a) and \ref{fig:lineprofileAB}(c)
 was assumed to represent the pseudo-diameter of the nanoparticle indicated by A--B in Fig.~\ref{fig:rba}.
By subtracting the estimated particle diameter from the pseudo-diameter of the nanoparticle,
 we obtained the value of twice the edge width.
The obtained edge widths were $6.9$ nm and $10.1$ nm for the nanoparticle in the HDE image and that in the output image, respectively.
Thus, the magnitude of the edge width in the output image is comparable to that in the HDE image.

\begin{figure*}
\centering
\includegraphics[width=\textwidth, bb = 40 150 900 390]{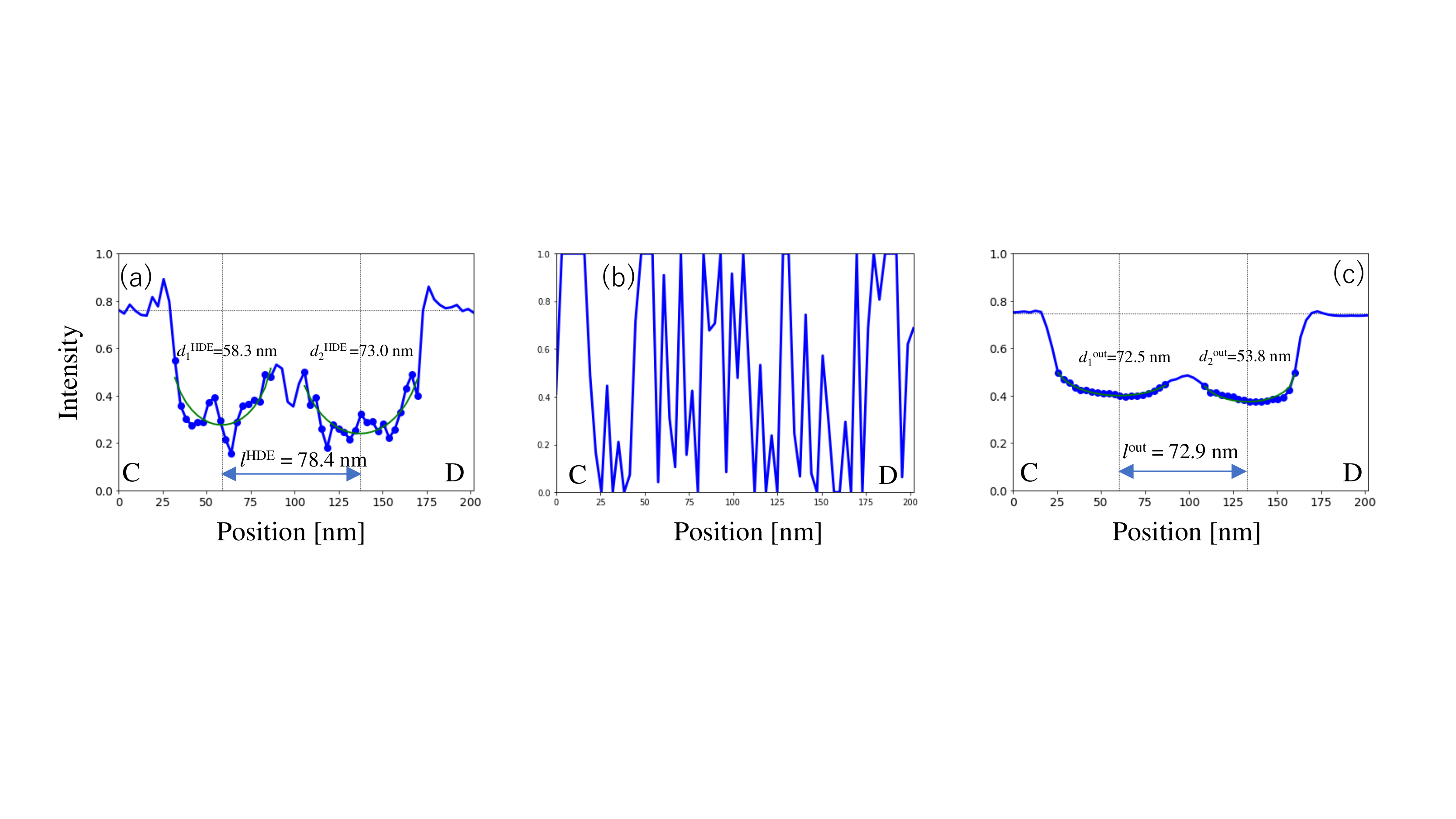}
\caption{
Line profile of C--D shown in Fig.~\ref{fig:rba}:
(a) the HDE image, (b) the LDE image, and (c) the output image.
Circles are the data used to calculate the diameter of the nanoparticle.
The dotted line shows the average of the background.
}
\label{fig:lineprofileCD}
\end{figure*}

We also investigated whether two adjacent nanoparticles could be distinguished.
The line profiles corresponding to C--D in Fig.~\ref{fig:rba} are shown in Fig.~\ref{fig:lineprofileCD}.
The line profile of the LDE image [Fig.~\ref{fig:lineprofileCD}(b)] shows random noise, as does that in Fig.~\ref{fig:lineprofileAB}(b).
At location C--D, nanoparticles are difficult to recognize in the LDE image.
By contrast, nanoparticles are recognized in the HDE and output images.
From the line profile of the HDE image [Fig.~\ref{fig:lineprofileCD}(a)],
 we obtained diameters of $d_{1}^{\rm HDE} = 58.3$ nm and $d^{\rm HDE}_{2} = 73.0$ nm for the nanoparticles in the HDE image.
From the nanoparticles' center position indicated by dotted vertical lines shown in Fig.~\ref{fig:lineprofileCD}(a),
 the distance between two nanoparticles is obtained as $l^{\rm HDE} = 78.4$ nm.
The condition $(d_{1}^{\rm HDE}+d^{\rm HDE}_{2})/2 < l^{\rm HDE}$ is satisfied; that is, the system has sufficient resolution to distinguish two nanoparticles.
In the case of the output image in Fig.~\ref{fig:lineprofileCD}(c),
 the sizes of nanoparticles are $d_{1}^{\rm out}=72.5$ nm and $d_{2}^{\rm out} = 53.8$ nm.
The difference in nanoparticle size determined from the HDE and output images is more than 20\%.
The distance between nanoparticles is $l^{\rm out} = 72.9$ nm, which is close to the $l^{\rm HDE}$.
The condition $(d_{1}^{\rm out}+d^{\rm out}_{2})/2 < l^{\rm out}$ is also satisfied for output image.
After applying our CNN model,
 we could distinguish between two adjacent nanoparticles at this size scale.

Our model removes noise from LDE images, clearly revealing the presence of nanoparticles.
The center position and the distance between adjacent nanoparticles can be reproduced at this size scale.
Although the size of the nanoparticles in the output image matches that in the HDE image,
 the nanoparticles' shape is uncertain (see Fig.~\ref{fig:rba}).
Moreover, the intensity of the line profile due to the internal structure of the nanoparticles can disappeared during the image processing
 because our model is a CNN that appears to spatially average the intensity.
Therefore, restoring the image of agglomerated nanoparticles like those in region I of Fig.~\ref{fig:rba},
 tends to be difficult, although isolated nanoparticles are correctly reproduced after our image processing.
For instance, we can count only five or seven nanoparticles in region I of the output image [Fig.\ref{fig:rba}(c)] against nine nanoparticles in the corresponding HDE image [Fig.~\ref{fig:rba}(a)].

\subsection{Waiting Time for Improvement}
\begin{table*}
\centering
\caption{
Waiting time for an output image in various processors.
The unit is in milliseconds.
The values were obtained as an average from 100 samples. 
}
\begin{tabular}{ccccc}
\hline
\begin{tabular}{c}\ \ \ Machine \ \ \ \\No.\end{tabular}  & Processor & \begin{tabular}{c}Model \\number\end{tabular} & Calc.& \begin{tabular}{l}Calc.\\including \\data transfer\end{tabular}\\
\hline
1 & CPU &  Intel  Core i9-9900X 3.50 GHz    & \rule[0em]{1em}{0em}$106 \pm 0.7$ \rule[0em]{1em}{0em}& --  \\
  & GPU &  NVIDIA Quadro RTX    8000        & \rule[0em]{1em}{0em}$4.4 \pm 0.0$ \rule[0em]{1em}{0em}& $8.0\pm 0.1$ \\
2 & CPU &  Intel  Core i7-9700  3.00 GHz    & \rule[0em]{1em}{0em}$300 \pm 10 $ \rule[0em]{1em}{0em}& --  \\
  & GPU &  NVIDIA Geforce GTX   1650        & \rule[0em]{1em}{0em}$5.3 \pm 0.6$ \rule[0em]{1em}{0em}& $25 \pm 0.5$  \\
\hline
\end{tabular}
\label{tab:wt}
\end{table*}

For \textit{in situ} TEM observations with low electron doses,
 improving the LDE image on a timescale that approaches the camera speed is important.
In maintain high performance of the TEM observation,
 the output image should be generated from the LDE image faster than the frame rate of the camera.
The framerates of the CMOS camera (OneView IS) are $25$ and $300$ frames per second in normal mode (output size: $4096 \times 4096$ pixels) and binning $8$ mode (output size: $512 \times 512$ pixels), respectively.
Our goal was to convert to one image within $40$ ms in the first step and then every $3.3$ ms thereafter.
Table \ref{tab:wt} shows the waiting time for converting an LDE image using our model and two types of machines.
The first configuration (Machine No. 1) is a calculation machine for numerical calculation,
 where the training in this study is performed.
The second configuration (Machine No. 2) is a personal computer for normal use.
In both cases, the CPU calculation requires hundreds of milliseconds.
The GPU calculation is fast, and the performance is more than 40 times greater than that of CPU calculation; that is, the conversion time is several milliseconds.
The time of conversion using the Geforce GTX 1650 graphics card is approximately the same as that using the Quadro RTX 8000 graphics card.
However, GPU calculation requires the data transfer from the CPU to the GPU and from the GPU to the CPU.
When the data transfer is taken into account, the total waiting time is 8 ms with Machine No. 1 and 25 ms with Machine No. 2.
Although the waiting time of 8 ms is more than two times longer than the 3.3 ms maximum temporal resolution of the Gatan OneView camera operating in binning 8 mode,
 it is substantially shorter than the 40 ms estimated from the frame rate of the OneView camera in normal use (25 fps).

\section{Summary}
We improved TEM images acquired using LDE by applying a simple CNN. 
Our model is based on the U-Net architecture with the ResNet encoder.
We demonstrated 
 that enabling the observation of objects that are difficult to visualize in the LDE image
 because our model can reproduce objects from the noise in addition to removing noise.
The position of nanoparticles in the HDE images was reproduced in the corresponding output images,
 and the size and the edge width of nanoparticles were similar in the HDE and output images.
In contrast,
 their shape reproduction requires further improvement.
The time necessary for the image conversion is approximately 8 ms, making the method applicable for \textit{in situ} observation at a frame rate of 125 fps or lower.
Our model is effective for investigating fast dynamic processes
 such as nucleation from a solution or tracking the motion of nanoparticles via LDE TEM observation.

\noindent\small\color{Maroon}\textbf{Acknowledgments }\color{Black}
This work is supported by JSPS KAKENHI Grant Numbers 20H05657 and 21K03379.


\begin{thebibliography}{999}
\providecommand{\urlprefix}{URL }


\bibitem[{Anada et~al., 2019}]{Anada-hologram}
 \textbf{Anada, S., Nomura, Y., Hiroyama, T. \& Yamamoto, K.} (2019).
 Sparse coding and dictionary learning for electron hologram denoising.
 \textit{Ultramicroscopy} \textbf{206}, 112818.

\bibitem[{Anderson et~al., 2013}]{Anderson-2013}
 \textbf{Anderson, H.S., llic-Helms, J., Rohrer, B., Wheeler, J., \& Larson, K.} (2013).
 Sparse imaging for fast electron microscopy.
 \textit{Proc. SPIE 8657, Computational Imaging XI} 86570C.


\bibitem[{Binev et~al., 2012}]{Binev-EMNST}
 \textbf{Binev, P., Dahmen, W., DeVore, R. \& Lamby, P.} (2012).
 Compressed sensing and electron microscopy,
 \textit{Modeling Nanoscale Imaging in Electron Microscopy}, Vogt, T., Dahmen, W. \& Binev, P. (Eds.), pp. 73--126. New York: Springer.


\bibitem[{Chen et~al., 2018}]{Chen-SID}
 \textbf{Chen, C., Chen, Q., Xu, J. \& Koltun, V. } (2018).
 Learning to See in the Dark.
 In \textit{2018 IEEE/CVF Conference on Computer Vision and Pattern Recognition}, pp. 3291--3300.


\bibitem[{Chlanda \& Sachse, 2014}]{Chlanda-rev-MMB2014}
 \textbf{Chlanda, P. \& Sachse, M.} (2014).
 Cryo-electron microscopy of viterous sections.
 \textit{Methods Mol Biol} \textbf{1117}, 193--214.


\bibitem[{Dabov et~al., 2007}]{Dabov-IEEE2007}
 \textbf{Dabov, K. Foi, A., Katkovnik, V. \& Egiazarian, K.} (2007).
 Image denoising by sparse 3-D transform-domain.
 \textit{2007 IEEE Trans on Image Process} \textbf{16}, 2080--2095.

\bibitem[{Danev \& Nagayama, 2001}]{Danev-UM2001}
 \textbf{Danev, R. \& Nagayama, K.} (2001).
 Transmission electron microscopy with Zernike phase plate.
 \textit{Ultramicroscopy} \textbf{88}, 243--252.


\bibitem[{Faruqi et~al., 2003}]{Faruqi-Ultramicrocopy2003}
 \textbf{Faruqi, A.R., Cattermole, D.M., Mikulec, B. \& Raeburn, C.} (2003).
 Evaluation of a hybrid pixel detector for electron microscopy.
 \textit{Ultramicroscopy} \textbf{94}, 263--276.


\bibitem[{Schneider et~al., 2014}]{Schneider-JCCP2014}
 \textbf{Schneider, N.M., Norton, M.M, Mendel. B.J., Gorgan, J.M., Ross, F.M. \& Bau H.H.} (2014).
 Electron-water interactions and implications for liquid cell electron microscopy.
 \textit{J Phys Chem C} \textbf{118}, 22373--22382.
 


\bibitem[{Gu et~al., 2014}]{Gu-IEEE2014}
 \textbf{Gu, S., Zhang, L., Zou, W. \& Feng, X.} (2014).
 Weighted nuclear norm minimization with application to image denoising.
 In \textit{2014 IEEE Conference on Computer Vision and Pattern Recognition}, pp. 2862--2869.


\bibitem[{Haider et~al., 1998}]{Haider-Nature1998}
 \textbf{Haider, M., Uhlemann, S., Schwan, E., Rose, H., \& Kabius, B.} (1998).
 Electron microscopy image enhanced.
 \textit{Nature} \textbf{392}, 768--769.


\bibitem[{He et~al., 2016}]{He-resnet}
 \textbf{He, K., Zhang, X., Ren, S., \& Sun, J.} (2016).
 Deep residual learning for image recognition.
 In \textit{2016 IEEE Conference on Computer Vision and Pattern Recognition}, pp. 770--778.



\bibitem[{Kingma \& Ba, 2015}]{Kingma-adam}
 \textbf{Kingma, D.P. \& Ba, J.} (2015).
 Adam: A Method for Stochastic Optimization.
In \textit{Proceedings of the 3rd International Conference on Learning};
 \textit{arXiv} 1412.6980v5.

\bibitem[{Kisielowski et~al., 2008}]{Kisielowski-2008}
 \textbf{Kisieloski, C., Breitag, B., Bischoff, M., van Lin, H., Lazar, S., Knippels, G., Tiemeijer, P., van der Stam, M., von Harrach, S., Stekelenburg, M., Haider, M., Uhlemann, S., M\"uller, H., Hartel, P., Kabius, B., Miller, D., Petrov, I., Olson, E.A., Donchev, T., Kenik, E.A., Lupini, A.R., Bentley, J., Pennycook, S.J., Anderson, I.M, Minor, A.M., Schmid, A.K., Duden, T., Radmilovic, V., Ramasse, Q.M., Vatanabe, M., Erni, R., Stach, E.A., Dense, P. \& Dahmen, U.} (2008).
 Detection of single atoms and buried defects in three dimensions by aberration-corrected electron microscope with 0.5-\AA information limit.
 \textit{Microsc Microanal} \textbf{14}, 469--477.


\bibitem[{Lim et~al., 2015}]{Lim-ICIP2015}
 \textbf{Lim, J., Kim, J.-H., Sim, J.-Y., \& Kim, C.-S.} (2015).
 Robust contrast enhancement of noisy low-light images: denoising-enhancement completion.
 \textit{2015 IEEE International Conference on Image Processing (ICIP)}, pp.4131--4135.


\bibitem[{Elad \& Aharon, 2006}]{Elad-IEEE2006}
 \textbf{Elad, M. \& Aharon, M.} (2006).
 Image denoising via sparse and redundant.
 \textit{2006 IEEE trans Image Process} \textbf{15}, 3736--3745.

\bibitem[{Morishita et~al., 2018}]{Morishita-MAM}
 \textbf{Morishita, S., Ishikawa, R., Kohno, Y., Sawada, H., Shibata, N. \&Ikuhara,Y.} (2018).
 Resolution achievement of 40.5 pm in scanning transmission electron microscopy using 300 kV microscope with delta corrector.
 \textit{Microsc. Microanal.} \textbf{24}, 120--121.

\bibitem[{Mor\'an \& Dahl, 1952}]{Moran-Science1952}
 \textbf{Fern\'andez-Mor\'an, H. \& Dahl, A.O.} (1952).
 Electron microscopy of ultrathin frozen sections of pollen grains.
 \textit{Science} \textbf{116}, 465--467.



\bibitem[{Ronneberger et~al., 2015}]{Ron-unet}
 \textbf{Ronneberger, O., Fischer P. \& Thomas, B.} (2015).
 U-Net: convolutional networks for biomedical image segmentation.
In Medical Image Computing and Computer-Assisted Intervention - MICCAI 2015., Lecture Notes in Computer Science , Vol. 9351.
Navab, N., Hornegger, J., Wells, W., Frangi, A. (Eds.), pp.234--241.
Springer, Cham.


\bibitem[{Shi et~al., 2016}]{Shi-arxiv2016}
 \textbf{Shi, W., Caballero, J., Hauzar, F., Totz, J., Aitken, A.P.,  Bishop, R., Rueckert, D. \& Wang, Z.} (2016).
 Real-time single image and video super-resolution using an efficient sub-pixel convolutional neural network.
 In \textit{2016 IEEE Conference on Computer Vision and Pattern Recognition}, pp. 1874--1883.

\bibitem[{Stevens et~al., 2014}]{Stevens-Mcro2014}
 \textbf{Stevens, A., Gang, H., Carin, L. ,  Arslan, I. \&  Brown, N.D.} (2014).
 The potential for Bayesian compressive sensing to significantly reduce electron dose in high-resolution STEM images.
 \textit{Microscopy} \textbf{63}, 41--51. 



\bibitem[{Yakubovskiy, 2020}]{smp-pytorch}
 \textbf{Yakubovskiy, P.} (2020).
 Segmentation Models: Python library with Neural Networks for Image Segmentation based on PyTorch. 
 GitHub repository, {\rm https://github.com/qubvel/segmentation\_models.pytorch}.







\end{thebibliography}
\end{document}